\documentclass[12pt]{iopart}  \usepackage{graphicx}
\usepackage{subfigure,cite} \usepackage{amsfonts,amssymb}

\newcommand{\text}[1]{\mbox{\scriptsize{#1}}}

\begin{document}

\title[Dynamics of Bacteriophage Genome Ejection]{Dynamics of
  Bacteriophage Genome Ejection {\it In Vitro\/} and {\it In
  Vivo}} 

\author{Debabrata Panja$^*$ and Ian J. Molineux$^{**}$}

\address{$^*$Institute for Theoretical Physics, Universiteit van
Amsterdam, Science Park 904,\\ Postbus 94485, 1090 GL Amsterdam, 
The Netherlands

$^{**}$ Molecular Genetics and Microbiology, Institute for Cell and
Molecular Biology, \\ The University of Texas, Austin, Texas 78712,
USA}

\begin{abstract} 
  Bacteriophages, phages for short, are viruses of bacteria. The
  majority of phages contain a double-stranded DNA genome packaged in
  a capsid at a density of $\sim500$ mg/ml. This high density requires
  substantial compression of the normal B form helix, leading to the
  conjecture that DNA in mature phage virions is under significant
  pressure, and that pressure is used to eject the DNA during
  infection. A large number of theoretical, computer simulation and
  {\it in vitro\/} experimental studies surrounding this conjecture
  has revealed many --- though often isolated and/or contradictory ---
  aspects of packaged DNA. This prompts us to present a unified view
  of the statistical physics and thermodynamics of DNA packaged in
  phage capsids. We argue that the DNA in a mature phage is in a
  (meta)stable state, wherein electrostatic self-repulsion is balanced
  by curvature stress due to confinement in the capsid. We show that
  in addition to the osmotic pressure associated with the packaged DNA
  and its counterions, there are four different pressures within the
  capsid: pressure on the DNA, hydrostatic pressure, the pressure
  experienced by the capsid, and the pressure associated with the
  chemical potential of DNA ejection. Significantly, we analyze the
  mechanism of force transmission in the packaged DNA, and demonstrate
  that the pressure on DNA is not important for ejection. We derive
  equations showing a strong hydrostatic pressure difference across
  the capsid shell. We propose that when a phage is triggered to eject
  by interaction with its receptor {\it in vitro}, the (thermodynamic)
  incentive of water molecules to enter the phage capsid flushes the
  DNA out of the capsid. {\it In vivo}, the difference between the
  osmotic pressures in the bacterial cell cytoplasm and the culture
  medium similarly results in a water flow that drags the DNA out of
  the capsid and into the bacterial cell.
\end{abstract}

\maketitle

\section{Introduction\label{sec1}}

Bacteriophages, or phages, are viruses of bacteria. Phages consist of
a protein capsid that encapsidates their genome, and a tail --- a
hollow tube connected to the capsid via a portal complex. During
infection, the phage tail attaches to a host bacterium, punctures the
cytoplasmic membrane and its genome translocates through the portal
and the tail into the bacterial cytoplasm. Infection initiates the
phage life-cycle: within the bacterial cytoplasm the genome is
transcribed and replicated, phage proteins are synthesized, and new
genome copies are packaged into newly assembled capsids. The cycle
ends with lysis of the host cell and the release of multiple
progeny. Understanding the mechanism(s) of phage genome ejection is
important, not only in the insights it provides on DNA structure but
also to provide a model for how eukaryotic viruses may release their
nucleic acid in to the cytoplasm or nucleus of an infected cell.

The genome of a mature phage virion is usually a B-form,
double-stranded DNA (dsDNA). A common phage may have 35-50 kb DNA (DNA
and genome will be used interchangeably) packed into a capsid of
$\sim60$ nm diameter, i.e., the phage DNA is packaged tightly within
the capsid in a condensed state, at a linear compression factor
$\sim250$, or at a density $\sim500$ mg/ml
\cite{north,earnshaw1,earnshaw2,cerritelli,hud1,olson,fokine,ari,chang,jiang,lander,xiang,tang,leiman,choi,comolli}. More
than five decades ago, this compression led to the conjecture
(henceforth referred to as the {\it ``pressure-conjecture''\/}) that
DNA in a mature phage capsid is under significant pressure, and is
kept in place by means of a ``plug'': protein(s) in the tail
tube. When the plug is opened by the action of the appropriate
receptor, the pressure of the DNA causes its release into the
cytoplasm of an infected cell in a biologically passive manner
\cite{hershey,stent,hayes,zarib}. The difficulty of obtaining
experimental kinetic data for the ejection of any phage genome into an
infected cell resulted in the conjecture becoming ``fact''.  The few
published examples that were inconsistent with the theory elicited
little response and most textbooks simply refer to the ``DNA
injection'' step without further elaboration. However, at the
beginning of this millennium, the pressure conjecture resurfaced in
the biophysics community. A single molecule study reported that the
packaging motor of phage $\phi29$ is capable of packaging DNA into a
phage capsid against a force of magnitude $\ge60$ pN \cite{smith}. The
force, which was assumed to be totally conserved, and thus available
for the subsequent ejection step, and which therefore can be termed as
the ``ejection force'', was simply defined as the pressure on the DNA
within the capsid, multiplied by the cross-sectional area of the unit
cell corresponding to the hexagonal lattice arrangement of the
packaged DNA. It was estimated that the pressure on the packaged DNA,
as well as the pressure on the inside of the capsid is in the order of
60 atm for a mature phage virion. This work was quickly followed by a
large number of computer simulations
\cite{locker,petrov1,petrov2,petrov3}, theoretical analyses
\cite{tzlil,purohit1,kindt,inam,purohit2}, and {\it in vitro\/}
experimental studies
\cite{evi1,gray3,evi2,jeem1,lef1,def,mang,gray1,rick,sao,evi3}
surrounding the pressure-conjecture, a significant fraction of these
studies stating the 60 atm pressure on the packaged DNA for a mature
virion as a matter of fact.

Many recent theoretical, computer simulation and {\it in vitro\/}
experi\-mental studies have been principally directed to quantify how
the pressure on the DNA can be understood from a thermodynamic
perspective. However, to the best of our knowledge, the issue of the
mechanical transmission of the pressure along the packaged DNA helix
has received no attention. Putting aside that question for the moment,
a general consensus has been reached on a thermodynamic description of
the pressure on the DNA. DNA is a charged polymer with persistence
length $\simeq50$ nm; DNA confined inside a capsid of diameter $\sim
50$ nm has a large free energy cost $F$, relative to the state of the
DNA outside the capsid. The pressure on the packaged DNA can be
thermodynamically derived from this free energy. However, fundamental
disagreements remain between different theories on the thermodynamic
origin of the pressure acting on the packaged DNA. In light of these
disagreements, together with the question of how mechanical
transmission of force along a flexible polymer, naturally raises the
question: ``Does the pressure-conjecture necessitate a biologically
passive ejection force''?

In order to appreciate the outstanding issues more easily, we provide
a brief overview of the structure of packaged phage DNA, and existing
approaches to describe phage DNA packaging and ejection.

\subsection{Structure of the packaged DNA in mature phage virions\label{sec1a}}

\begin{figure}[h]
\begin{center}
\includegraphics[width=0.8\linewidth]{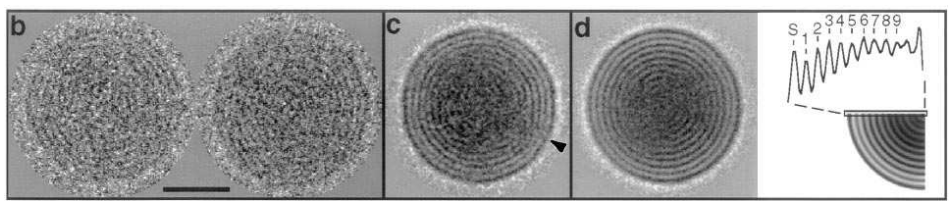}
\end{center}
\caption{Source: Ref. \cite{cerritelli}. (b) Cryo-electron images of
  mature T7 tail-deletion mutant virions, viewed along the portal
  axis; bar corresponds to 25 nm. (c) Image obtained by averaging over
  21 virions, the closed triangle marks the discontinuity between the
  second and the third DNA-associated rings. (d) Averaging over 77
  virions; on the right appears the azimuthally averaged image: S is a
  dense ring corresponding to the capsid, and 9 other peaks
  corresponding to 9 DNA rings. Reproduced with permission from
  Elsevier Inc. \label{fig1}}
\end{figure}

\begin{figure}[h]
\begin{center}
\begin{minipage}{0.18\linewidth}
\includegraphics[width=\linewidth]{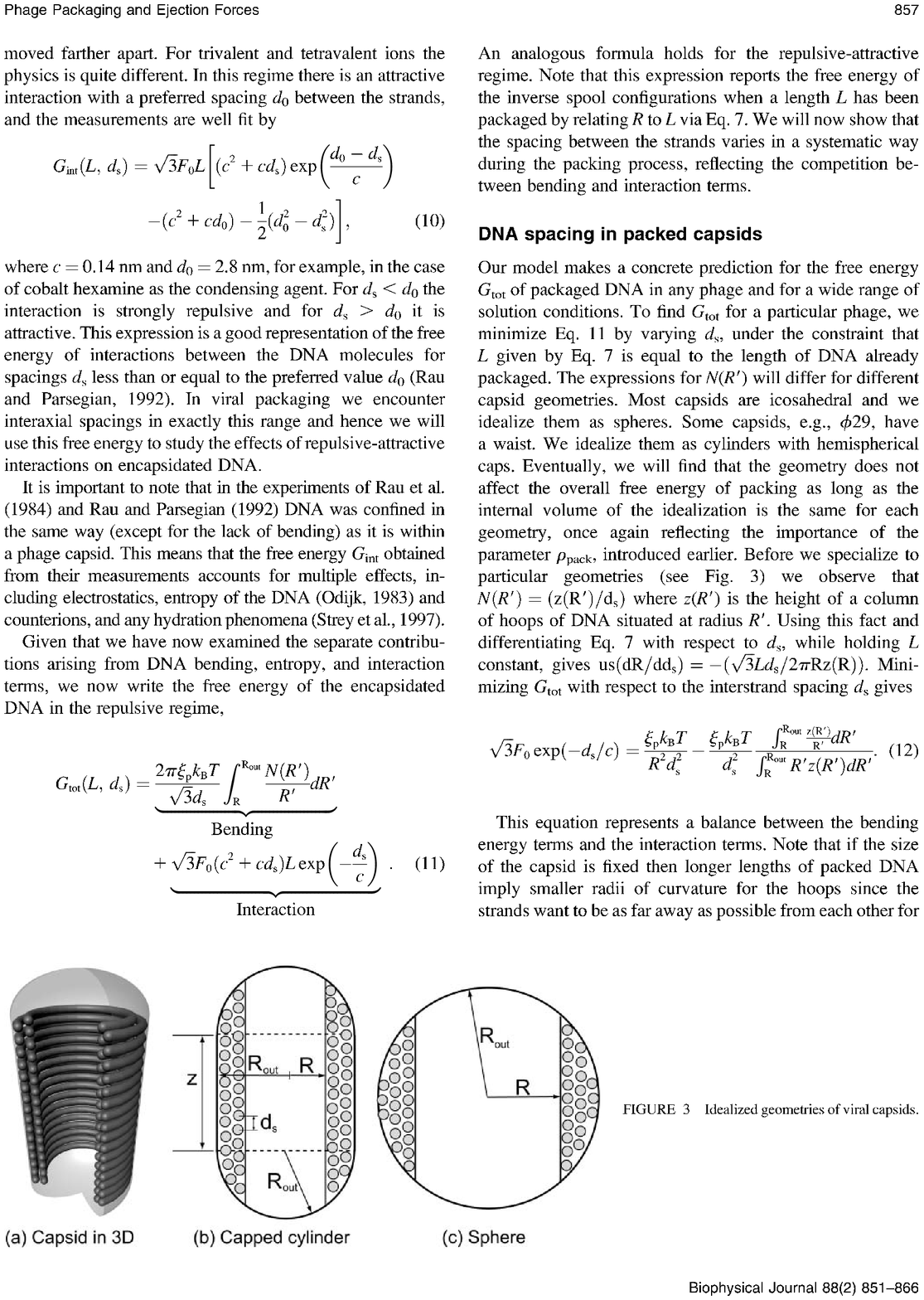}
\end{minipage}
\hspace{2cm}
\begin{minipage}{0.5\linewidth}
\includegraphics[width=\linewidth]{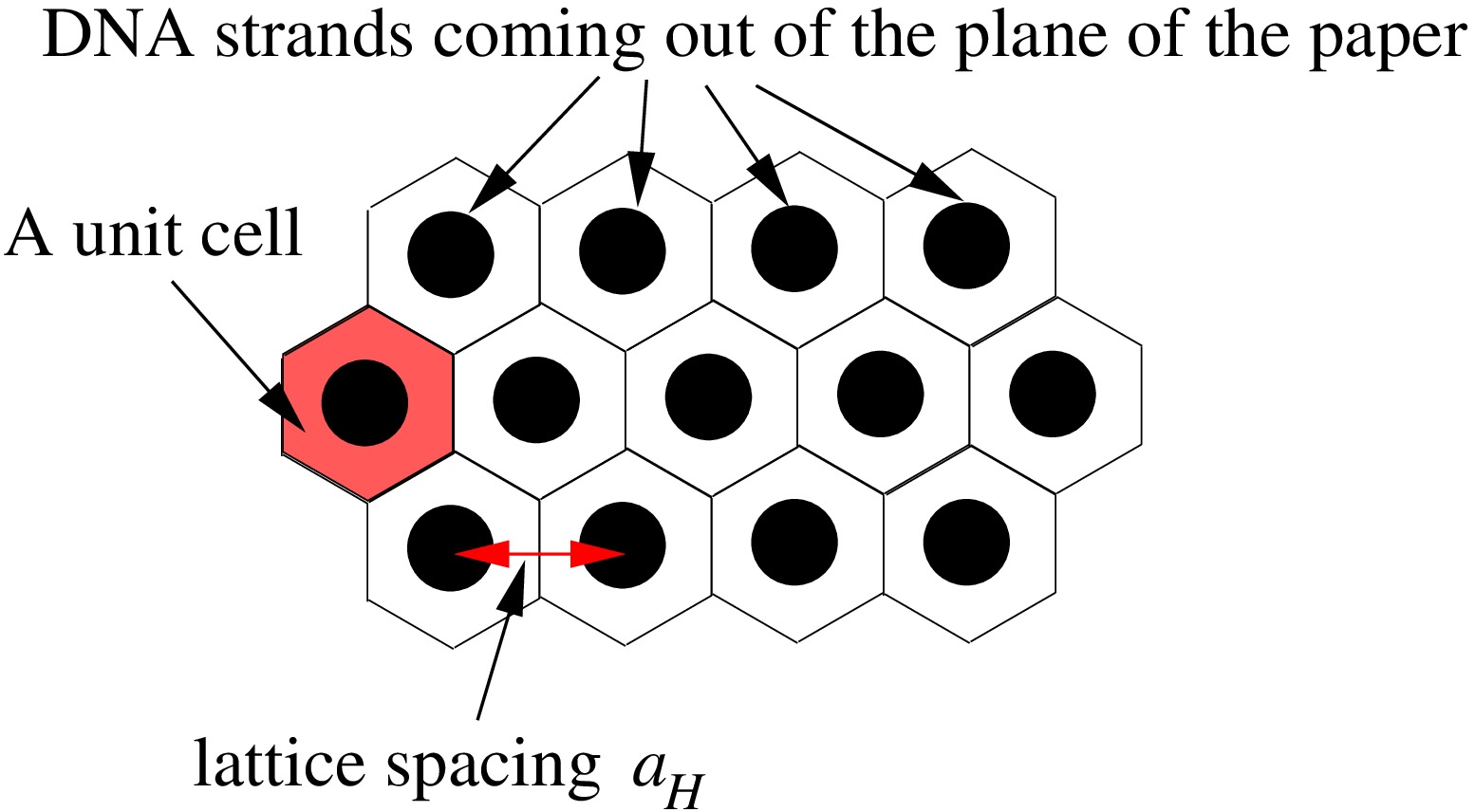}
\end{minipage}
\end{center}
\caption{(Left) Source: Ref. \cite{purohit1}.  Idealized DNA packaging
  in a coaxial toroidal form within the capsid. Reproduced with
  permission from Elsevier Inc. (Right) Hexagonal lattice formation by
  the DNA spool, within the capsid. The black filled circles at the
  center of each hexagonal unit cell (an example is painted pinkish
  red) denotes the cross-section of the DNA.\label{fig2}}
\end{figure}

Experiments on the structure of the packaged DNA within phage capsids
have a long history: preliminary evidence of hexagonal packaging of
the DNA for T2 and T7 phages was first obtained in 1961 \cite{north},
and X-ray diffraction was also used to establish that DNA forms
concentric layers within the capsid
\cite{earnshaw1,earnshaw2,cerritelli}. Several theoretical models have
been proposed for the topology of the packaged DNA: wound into a spool
\cite{riemer,harri,gabash}, a liquid crystal with hairpins
\cite{black,ser} or defects \cite{lep}, to name but a few, subjecting
the topic to much debate. The first unambiguous high-quality images of
DNA organization in mature phage capsids were obtained using
cryo-electron microscopy in 1997. Averaged over many mature T7
tail-deletion mutant virions, viewed along the portal axis, the
packaged DNA showed patterns of circular striations, spaced
$\approx2.5$ nm apart (Fig. \ref{fig1}) \cite{cerritelli}. A
computer-modeled projection (side views) of spooled DNA within the
capsid, again averaged over many virions, strongly suggested that T7
DNA is wrapped axially in concentric shells of a toroid. In each
shell, DNA is coiled with an axial rise of $\approx2.5$ nm per turn,
and the diameter of each turn is imposed by the previous
shell. Following this pioneering study, a succession of cryo-electron
microscopy reconstructions have provided support for a toroidal
structure: $\lambda$ \cite{hud1}, both isometric and prolate T4
\cite{olson,fokine}, P22, $\epsilon15$, $\phi29$
\cite{ari,chang,jiang,lander,xiang,tang}, K1E and K1-5 \cite{leiman},
P-SPP7 \cite{liu} and N4 \cite{choi} all reveal consecutive layers of
toroidal DNA spaced by $\approx2.5$ nm. Together, these observations
suggest that, when averaged over many virions, the well-defined
hexagonal lattice for the packaged DNA may be a generic feature of all
phages (Fig. \ref{fig2}). Separately, it should be noted that, where
it was resolved in these reconstructions, the leading end (first end
to enter the infected cell) is seen to extend into the narrow portal
channel.  Furthermore, in earlier studies, the leading end of the
genome could be cross-linked to the tail of several mature phage
virions \cite{chatto,saigo1,saigo2,thomas}.  Portal/tail insertion of
the leading DNA end during virion morphogenesis is also likely a
general feature as it ensures that the subsequent DNA ejection step is
efficient.

It is of paramount importance to emphasize that the hexagonal toroidal
spool structure of the packaged DNA is an {\it average\/}
property. This point is illustrated beautifully in a recent paper on
phage T5 \cite{lef2}.  Further, to be packed within a capsid the DNA
helix has to cross itself and therefore cannot lie on a perfect
lattice. The toroidal structure has also been suggested to hold only
for the part of the DNA close to the inner wall of the capsid, and not
for the entire DNA (e.g., Refs. \cite{comolli,petrov2}).

\subsection{Existing approaches for the thermodynamics of packaging
  and ejection forces\label{sec1b}}

There have been two main --- and mutually conflicting --- approaches
that address the thermodynamic origin of packaging and ejection
forces: continuum mechanics and coarse-grained molecular mechanics
models. In the continuum mechanics model
\cite{tzlil,purohit1,kindt,inam,purohit2}, packaged DNA is seen as a
uniformly charged rod, with a certain persistent length $l_p$,
organized in a toroidal spool on a hexagonal lattice
(Fig. \ref{fig2}). The continuum mechanics model is simple enough to
allow analytical calculations: packaged DNA is assumed to repulsively
self-interact (with electrostatic Coulomb interaction), allowing
calculation of the stored electrostatic energy $U^{\text{(el)}}_{\text
  p}$. Similarly, the bending energy $U^{\text{(b)}}_{\text p}$ is
obtained from the toroidal spool arrangement of DNA on a hexagonal
lattice. Both the electrostatic and bending energies are assumed to be
zero for free DNA outside the capsid. Thereafter, with the assumption
that the difference between the entropy of free and packaged DNA is
negligible, the difference between the Helmholtz free energy of
packaged and free DNA is obtained as $F = U^{\text{(el)}}_{\text p} +
U^{\text{(b)}}_{\text p}$. Note that in a given buffered solution and 
a capsid of volume $V$, made of a perfectly rigid material, $F$ is
only a function of the length $L$ of the DNA within the capsid, i.e.,
$F\equiv F(L)$. In particular, $F$ {\it increases\/} as a function of
$L$, and it is $-\partial F/(A_{\text{DNA}}\partial L)$, the pressure
associated with the chemical potential of DNA ejection (the chemical
potential of DNA ejection is simply defined as the negative rate of
change of the free energy $F$ with respect to the length of DNA within
the capsid) that manifests as the ejection pressure, where
$A_{\text{DNA}}$ is the cross sectional area of the DNA. Note that
$\partial F/\partial L$ has the units of force; consequently, in the
continuum mechanics model, $-\partial F/\partial L$ is viewed as the
ejection force, and $\partial F/\partial L$ as the packaging force
(note that $-\partial F/\partial L$ is also the chemical potential of
DNA ejection, as defined above).

In the {\it in vitro\/} $\lambda$ experiments
\cite{evi1,gray3,evi2,jeem1}, phage virions are immersed in a solution
containing PEG and/or DNA condensing agents, and DNA is ejected when
triggered by the LamB receptor protein. The length of the DNA that
remains within the capsid at the end of the ejection process depends
on the concentration of the PEG and/or DNA condensing agent. (The
basic principles underlying these {\it in vitro\/} experiments were
first presented in a theoretical paper more than four decades ago
\cite{zarib}). The continuum mechanics model has been widely
advertised to explain the {\it in vitro\/} ejection data, not just for
$\lambda$ but for all phages --- both {\it in vitro\/} and {\it in
  vivo}. A somewhat deeper look, however, reveals that the agreement
between theory and the {\it in vitro\/} experiments using T5 (one of
only three phages studied in osmotic suppression experiments), remains
poor \cite{lef1,def}. The model also neglects how the ejection of
proteins from infecting virions (which is a feature of most phages)
into the cell is accomplished, and it cannot therefore describe the
complete infection process.  Further, there is also disagreement
between the continuum mechanics theory and the experimental kinetics
of T7 DNA translocation into the bacterial cytoplasm {\it in vivo\/}
\cite{kemp,mol3}. All but the leading 0.5 kb of the 70 kb genome of
the unrelated phage N4 is also known to be internalized by
transcription in the cell \cite{choi,kazm}, and marker rescue
experiments with phage SP82 suggest that the phage genome enters a
cell at a constant rate \cite{mcallister}.  None of the {\it in
  vivo\/} experiments that show kinetics of genome internalization are
consistent with theories derived from {\it in vitro\/} ejection
studies.

The continuum mechanics model suffers from fundamentally unrealistic
assumptions that diminish its value. In particular, {\it free energy
  is an equilibrium concept}; obtaining an ejection pressure by
$-\partial F/(A_{\text{DNA}}\partial L)$, in order to explain the
experimental data on the irreversible ejection process, has the
underlying assumption that DNA remains at equilibrium in a toroidal
spool configuration on a hexagonal lattice {\it at all stages of
  ejection}. Not only is this assumption unsubstantiated by polymer
physics theory, it is also inconsistent with actual experimental data
on T5 \cite{lef2}; the latter is illustrated in Fig. \ref{fig2a}
below. In order to circumvent some of these problems, the continuum
mechanics model invokes two free parameters to fit experimental data.
\begin{figure}[h]
\begin{center}
\includegraphics[width=\linewidth]{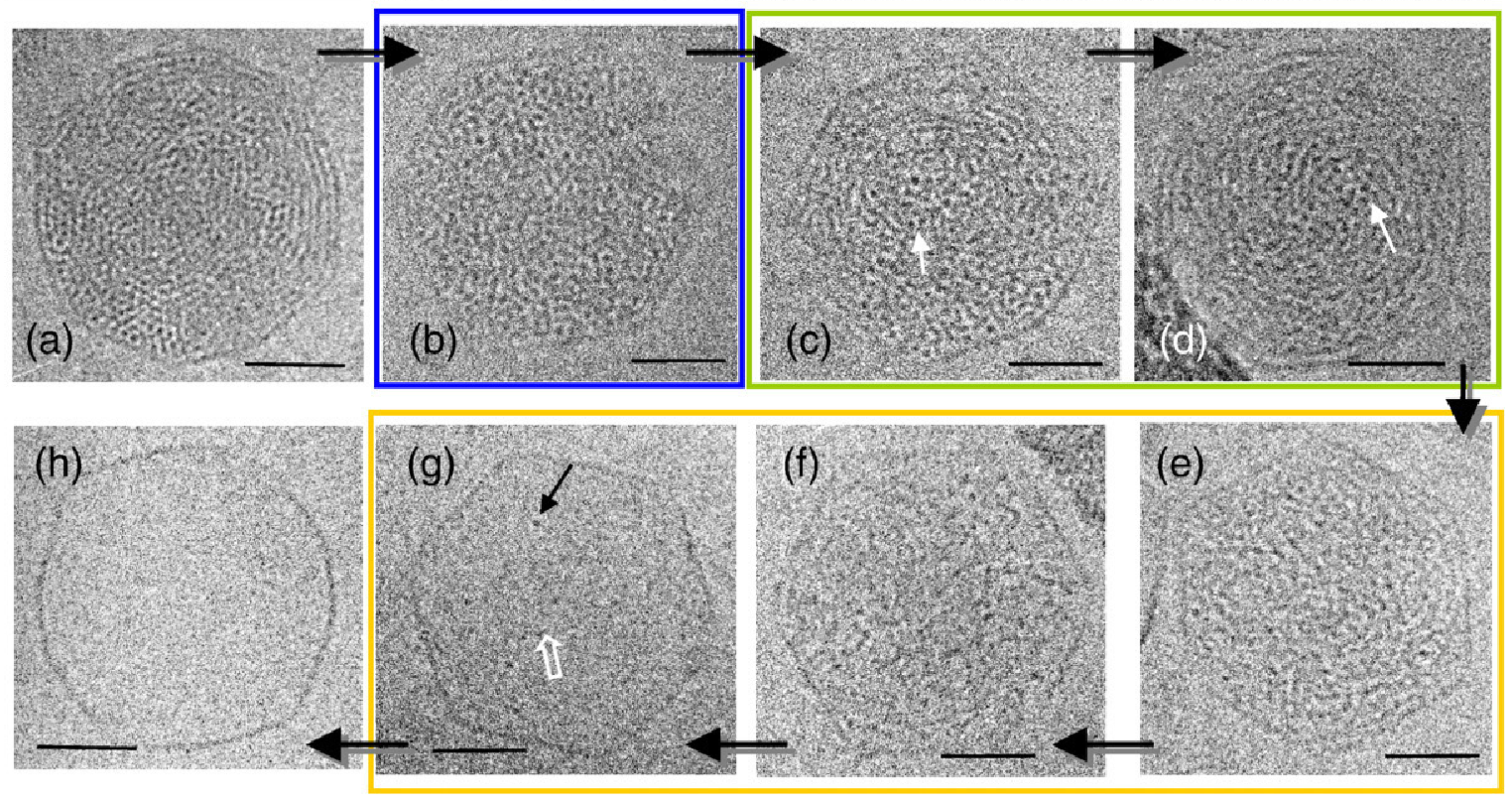}
\end{center}
\caption{Cryo-EM images of a single T5 virion at different stages of
  ejection: (a) full capsid to (h) empty capsid, showing the {\it
    lack\/} of toroidal spool configuration of the DNA hexagonal
  lattice at all stages of ejection, in contrast to the assumption
  made by the continuum mechanics model. Figure reproduced from
  Ref. \cite{lef2}, with permission from Elsevier Inc.\label{fig2a}}
\end{figure}

In coarse-grained molecular mechanics models
\cite{locker,petrov1,petrov2,petrov3}, DNA is modeled as a polymer,
with a persistence length $l_p$, and polymer dynamics is allowed to
take its own course in simulations of phage genome packaging and
ejection. The model does not allow analytical calculations --- it is
effected only by computer simulations, and in work published to date,
the DNA is modeled without surrounding water and counterions. However,
there is no assumption that DNA remains at equilibrium in a toroidal
spool configuration on a hexagonal lattice during packaging or
ejection. The main limitation of coarse-grained molecular mechanics
model is that because processes are only materialized by computer
simulation, they are difficult to correlate with experimental data.

\subsection{This paper\label{sec1c}}

In summary, our understanding of the thermodynamics of phage DNA
packaging and ejection, provided by the existing literature, is far
from satisfactory. In this paper we present a comprehensive treatise
of the thermodynamics of packaged DNA and the mechanism of genome
ejection, {\it independent of any specific model}. In Sec. \ref{sec2}
we show that there are four distinct pressures within the capsid: 
pressure on DNA, hydrostatic pressure, pressure experienced by the 
capsid, and pressure associated with the chemical potential of DNA 
ejection. {\it These are all different thermodynamic quantities.} We 
also show that for amature phage the hydrostatic pressure within the 
capsid is much higher than outside, and that the pressure on the DNA 
is a lot smaller than commonly envisaged. In Sec. \ref{sec3} we take 
up the issue of force transmission along the DNA helix, and show that 
the pressure on DNA is not important for genome ejection. Instead, 
in Sec. \ref{sec4}, we suggest that when a phage is triggered {\it 
in vitro\/} to eject by interaction with its receptor, the 
(thermodynamic) incentive for water to enter the capsid flushes the 
DNA out. We argue that {\it in vivo}, the difference between the 
osmotic pressure within the cell cytoplasm and the outside culture 
medium initiates, and {\it importantly maintains}, a flow of water 
from the culture medium into the capsid, and then from the capsid 
down the phage tail-tube into the bacterial cell cytoplasm, 
dragging the DNA out of the capsid and into the cell. We show that 
this theory is consistent with both studies of {\it in vitro\/} 
ejection and observations of infection {\it in vivo}.

\section{Thermodynamics of phages {\emph{in vitro\/}}\label{sec2}}

\subsection{Entropy of DNA ejection {\it in vitro\/}\label{sec2a}}

We consider an experiment with an ensemble of realizations, each
containing a phage capsid initially placed in a certain buffer
solution: for each realization a DNA of length $L$ is packaged within
a capsid, whose volume we denote by $V$. The DNA is allowed to exit
the phage {\it in vitro}, and at the end of the experiment the entire
genome ends up in the buffer. The experimental system is kept at a
temperature $T$ at all times. The only interaction of the system with
the environment is to exchange (thermal) energy. For the system in its
entirety, we define the initial (DNA within the capsid) free-energy
$F_{\text p}$ and the entropy $S_{\text p}$; the final (DNA in the
buffer) free-energy $F_{\text{unp}}$ and entropy $S_{\text{unp}}$. The
quantities $F_{\text{ejection}} = F_{\text{unp}} - F_{\text p}$, and
$S_{\text{ejection}} = S_{\text{unp}} - S_{\text p}$ are then the free
energy and entropy of ejection, respectively.

The above isothermal experiment, when actually performed in a
laboratory, is exothermic \cite{jeem2}. Earlier experiments using
differential scanning calorimetry led to the same conclusion
\cite{mdzin,ser1,ser2}; i.e., the system releases an amount of heat,
of magnitude $Q_{\text{ejection}}$ to the environment. This tells us
that $S_{\text{ejection}} < 0$; however, the amount of heat released
to the environment cannot give us the value $S_{\text{ejection}}$
because the ejection process is spontaneous and irreversible. All we
can say, using the second law of thermodynamics, is that
$|S_{\text{ejection}}| \ge Q_{\text{ejection}}/T$.

The result that the entropy of ejection is negative, from the point of
view of DNA confinement in the capsid, is counter-intuitive. One would
expect the configurational entropy of the DNA to be strongly reduced
due to confinement when compared to the same DNA in solution; in other
words, from the configuration entropy of the DNA in isolation one
would expect $S_{\text{ejection}} > 0$. This is, in fact, correct;
however, the DNA cannot be considered in isolation. The
configurational entropy of the DNA is only a very small part of the
entropy of the entire system; most of the change in entropy associated
with ejection of DNA from its confined state within the capsid into
the environment involves the ordering of water molecules around the
free DNA \cite{leikin}. DNA, being a charged molecule, orders the
dipole orientation of nearby water molecules. There are many fewer
water molecules surrounding the DNA inside the capsid when compared to
the same DNA in solution. Consequently, although the DNA molecule does
gain entropy following ejection from the capsid, so many more water
molecules lose entropy that the entropy of the entire system decreases
as DNA is ejected from the capsid.

This issue can be used to evaluate the current approaches for
analyzing the physics of phage ejection. DNA configurational entropy
is not considered a significant parameter in the continuum mechanics
model, while molecular mechanics models
\cite{locker,petrov1,petrov2,petrov3} show that the configurational
entropy constitutes a sizable fraction of the free energy of the
packaged DNA.

\subsection{Thermodynamics for the capsid content of a mature phage in
  a buffer solution\label{sec2b}}

Most phage capsid shells are fully permeable to water and small ions
or molecules, but are impermeable to large molecules in the buffer or
to the encapsidated DNA. Some phage capsids, like those of T4 and its
relatives, are much less permeable, even to small molecules or ions,
including cesium chloride and ammonium acetate.  Such capsids can
easily be broken by osmotic shock, a phenomenon that resulted in the
classic electron micrograph showing phage T2 DNA outside a ruptured
virion \cite{klein}.

We assume that the entire system, with DNA packaged within the capsid,
can be thermodynamically separated into two subsystems separated by
the capsid shell. The subsystem within the capsid consists of DNA
(perhaps in addition to protein molecules, which we will not
specifically refer to further), and an aqueous environment containing
small solutes (including ions) that can permeate the capsid shell,
while the subsystem outside usually consists of water and solute
molecules that can or cannot permeate the capsid shell. Apart from
exchanges of solutes that can permeate the capsid shell, these two
subsystems do not interact. This assumption allows us to express
$F_{\text p} = F_{\text{p,in}} + F_{\text{p,out}}$, where
$F_{\text{p,in}}$ and $F_{\text{p,out}}$ are, respectively, the free
energy of the contents of the capsid and that of the environment
outside the capsid. Clearly, $F_{\text{p,in}}$ is a function of the
equilibrium capsid volume $V$, DNA length $L$, number of water
molecules $\{n_{\text{w,in}}\}$ and the number of solute molecules of
all species $\{n_{\text{s,in}}\}$ within the capsid. Similarly,
$F_{\text{p,out}}$ is a function of number of water molecules
$\{n_{\text{w,out}}\}$ and the number of solute molecules of all
species $\{n_{\text{s,out}}\}$ in the solution. Both subsystems are
``open'' in the thermodynamic sense, since they exchange water and
small solute molecules. We assume (realistically) that the volume of
the buffer in which the phage is immersed is $\gg V$, so for analyzing
the thermodynamics of the phage and its contents, the external
solution can be considered to be an infinite reservoir for water and
small permeable solute molecules. The number of solute molecules
within the phage capsid $\{n_{\text{s,in}}\}$ is then fixed by the
chemical potential of the small solute molecules in the buffer,
collectively denoted by $\{\mu_{\text s}\}$ (for the case of
counterions, via the Donnan equilibrium). This simply means that
$F_{\text{p,in}}\equiv F_{\text{p,in}}(V,L)|_{\{\mu_{s}\}}$. The free
energy $F_{\text{p,in}}(V,L)|_{\{\mu_{\text s}\}}$ can be further
dissociated into the interaction free energy
$F^{\text{(int)}}(V,L)|_{\{\mu_{\text s}\}}$ of the ions, DNA and
water molecules within the capsid, and the DNA configurational free
energy (this includes bending, or curvature, energy and the
configurational entropy of the DNA) $F_c(V,L)|_{\{\mu_{\text s}\}}$ of
the DNA is then given by
\begin{eqnarray}
  F_{\text{p,in}}(V,L)|_{\{\mu_{\text
      s}\}}=F_{\text{p}}^{\text{(int)}}(V,L)|_{\{\mu_{\text s}\}}+F_c(V,L)|_{\{\mu_{\text s}\}}.
\label{e0}
\end{eqnarray} 
We also assume that water is incompressible and that small solutes do
not occupy any physical volume. With these assumptions, in
equilibrium, the number of water molecules within the capsid is then
determined by $V$ minus the physical volume of the DNA, and we can now
explicitly focus on a number of thermodynamic issues.

\subsubsection{Pressure associated with the chemical potential for DNA
  ejection\label{sec2b2}}

If we assume that the capsid is made of a perfectly rigid material,
then from the above definitions
$F_{\text{p,in}}(V,L)|_{\{\mu_{s}\}}\equiv
F_{\text{p,in}}(L)|_{\{\mu_{s}\}}$, we can define the pressure
associated with the chemical potential for DNA ejection. Consider two
situations using the same buffer conditions (i.e., at fixed chemical
potentials of permeable ions), one where DNA of length $L$ and the
other a length of $L-dL$, is packaged within the capsid. We assume
that Donnan equilibrium conditions are satisfied for both lengths of
DNA within the phage capsid. One can then define the pressure
associated with the chemical potential of DNA ejection by comparing
the free energy of the capsid content under these two situations,
viz. $F_{\text{p,in}}(L)|_{\{\mu_{s}\}}$ and
$F_{\text{p,in}}(L-dL)|_{\{\mu_{s}\}}$, as:
\begin{eqnarray}
  P_{\text{ejection}}=-\frac{1}{A_{\text{DNA}}}\frac{\partial
    F_{\text{p,in}}(L)|_{\{\mu_{s}\}}}{\partial L},
\label{e4}
\end{eqnarray}
where $A_{\text{DNA}}$ is the cross-sectional area of the DNA. The
pressure $P_{\text{ejection}}$ is the pressure that the plug at the
end of the tail tube feels from within the phage.

\subsection{Pressure on the capsid from within, hydrostatic pressure
  imbalance across the capsid and pressure on the DNA\label{sec2c}}

In Sec. \ref{sec2b}, we considered the thermodynamics of the capsid
content as a subsystem, under a fixed chemical potential of permeable
small solute molecules in the buffer, while assuming that the number
of water molecules within the capsid is simply determined by the
capsid volume. This is however an incomplete description of the
thermodynamics of the entire system (namely the phage immersed in the
buffer solution), as we did not consider the thermodynamic penalties
associated with exchange of water molecules across the capsid
shell. If the capsid is made of a perfectly rigid material without the
possibility of expansion or contraction, such a description would be
correct, but in reality, the capsid is comprised of protein molecules,
and therefore has a finite, albeit high, rigidity \cite{iva1}. Keeping
the assumption that water is incompressible, rather than a fixed
capsid volume determining the number of water molecules internalized,
the capsid volume should actually be determined by the thermodynamics
of water molecule exchange across the capsid shell, which is
equivalent to a semi-permeable membrane.  

We now argue that for a given length $L$ of the DNA within the capsid, 
the equilibrium volume $V$ is determined by the buffer composition. We
presuppose that an expansion/contraction of the capsid allows for an
exchange of a small number of water molecules, relative to the total
number of water molecules present in the external buffer. We can then
safely assume that the chemical potential of permeable small solutes
in the buffer remains unaltered by the exchange of water molecules
across the capsid shell.

First, the pressure on the capsid from within, $P_{\text{capsid}}$,
can be calculated from the $V$ -dependence of the free energy
$F_{\text{p,in}}(V,L)|_{\{\mu_{s}\}}$ as follows. Imagine a virtual,
uniform expansion of the capsid leading to a volume increase from $V$
to $V+dV$ (the extra volume will be filled by water from the buffer
and permeable solute molecules as dictated by $\{\mu_{\text s}\}$ ---
for the case of ions, via the Donnan equilibrium). For a given length
of DNA, under fixed $\{\mu_{\text s}\}$, this leads to a new value of
the free energy of the inside material of the capsid,
$F_{\text{p,in}}(V+dV,L)|_{\{\mu_{s}\}}$, but also, since the buffer
solution loses a volume $dV$ of water, the free energy of the buffer
solution increases by an amount $\pi_{\text{out}} dV$, where
$\pi_{\text{out}}$ is the osmotic pressure of the buffer solution. The
pressure on the capsid from within, is therefore given by
\begin{eqnarray}
P_{\text{capsid}}=-\frac{\partial
  F_{\text{p,in}}(V,L)|_{\{\mu_{s}\}}}{\partial V}-\pi_{\text{out}}.
\label{e3}
\end{eqnarray}

Secondly, water's thermodynamic incentive to enter the capsid is given
by the osmotic gradient across the capsid shell, namely
$[\pi_{\text{in}}-\pi_{\text{out}}]$, where $\pi_{\text{in}}$ is the
osmotic pressure within the capsid shell, and is defined by [see
Eq. (\ref{e0})]
\begin{eqnarray}
  \pi_{\text{in}}=-\frac{\partial
    F_{\text{p}}^{\text{(int)}}(V,L)|_{\{\mu_{s}\}}}{\partial V}.
\label{e3a}
\end{eqnarray}
Simultaneously, when entering the capsid, the water molecules exit a
zone of hydrostatic pressure $P_{\text{hydro,out}}$, the hydrostatic
pressure in the solution outside the capsid, and enter the capsid
which may have, in principle, a different pressure,
$P_{\text{hydro,in}}$. The equilibrium capsid volume is then
determined by the condition that
$\left[\pi_{\text{in}}-\pi_{\text{out}}\right]$ is counter-balanced by
the pressure-volume work done by water molecules in entering the
capsid
\begin{eqnarray}
  P_{\text{hydro,in}}-P_{\text{hydro,out}}=\pi_{\text{in}}-\pi_{\text{out}}.
\label{e6}
\end{eqnarray}
It is interesting to note that Eq. (\ref{e6}) is well-known to hold
for bacterial and plant cells: the osmotic pressure gradient is
counter-balanced by a hydrostatic pressure differential (turgor)
\cite{koch}. Turgor allows cells to enlarge and thus facilitates
growth.

Further, an appreciation for Eq. (\ref{e6}) is afforded by a
straightforward gedanken experiment that connects to elementary
physical chemistry. Consider a vertical U-tube with a membrane at the
lowest point separating the two arms. The membrane is permeable to
water, but impermeable to, say, sugar molecules. We fill up the two
arms of the U-tube to equal height: the left arm with pure water, and
the right arm with sugar solution. As time progresses, water from the
left arm will permeate into the right arm, reducing the height of the
water column on the left, while increasing the height of the sugar
solution column on the right. At equilibrium, the height of water in
the left column will be lower than the sugar solution, meaning that
there is a hydrostatic pressure difference $P$ across the
membrane. Moreover, at equilibrium the concentration of sugar in the
right column is non-zero, while it remains zero in the left column by
construction. There is therefore an osmotic gradient across the
membrane, which is precisely counter-balanced by the hydrostatic
pressure $P$. Equation (\ref{e6}) describes the same equilibrium,
where the difference between the osmotic pressures inside and outside
the capsid is counter-balanced by the hydrostatic pressure difference
across its protein shell.

Equations (\ref{e3}) and (\ref{e6}) allow us to appreciate not only
how the pressure on the capsid of finite rigidity is mechanically
materialized, but also how the pressure $P_{\text{DNA}}$ that the
capsid mechanically transmits to the DNA (and by action-reaction, the
pressure the DNA transmits to the capsid). First, the force balance on
a small capsid surface tells us that the capsid pressure
$P_{\text{capsid}}$ equals the hydrostatic pressure gradient across
the capsid plus the pressure that is mechanically transmitted to the
DNA inside; i.e.,
\begin{eqnarray}
P_{\text{capsid}}=P_{\text{hydro,in}}-P_{\text{hydro,out}}+P_{\text{DNA}}.
\label{e5}
\end{eqnarray}
Incorporating Eqs. (\ref{e0}) and (\ref{e3a}-\ref{e5}), this reads
\begin{eqnarray}
  P_{\text{DNA}}=-\frac{\partial
    F_{\text{p,in}}(V,L)|_{\{\mu_{s}\},L}}{\partial V}-\pi_{\text{in}}=-\frac{\partial
    F_c(V,L)|_{\{\mu_{s}\}}}{\partial V}.
\label{e7}
\end{eqnarray}
Note in Eq. (\ref{e7}) that the osmotic pressure within the capsid
$\pi_{\text{in}}$ is derived from the interaction energy of the ions,
DNA and water molecules within the capsid, while the same interaction
energy appears in $F_{\text{p,in}}(V,L)|_{\{\mu_{s}\}}$, but with
opposite sign [Eq. (\ref{e3a})]. Consequently, the net contribution of
this interaction energy to the r.h.s. of Eq. (\ref{e7}) is zero. This
conclusion implies that the pressure on the DNA mechanically
transmitted by the capsid comes entirely from the stiffness of the
DNA. We verify this conclusion in Sec. \ref{sec3} for a single turn of
the DNA. Therein we also discuss why Eq. (\ref{e7}) is not in conflict
with the experiment by Smith {\it et al.} \cite{smith}.

It is straightforward to show that Eqs. (\ref{e3}-\ref{e7}) are
consistent with the Gibbs free energy minimization of the entire
system (the mature phage plus the buffer solution). Moreover, we note
that due to the cancellation of the interaction energy term in Eq. (7)
as explained above, the compressive forces on the DNA mechanically
transmitted by the capsid are small compared to $P_{\text{capsid}}$:
the bulk of $P_{\text{capsid}}$ is mechanically transmitted to the
water inside. This is in fact confirmed from the structural and
interaction properties of both the packaged DNA and the structure of
the protein molecules comprising the capsid shell.  Raman spectral
studies of mature P22 and T7 phages have not found any evidence of any
structural alteration of protein or DNA that would be expected if they
were subjected to high pressure \cite{george1,george2}. Only the
configuration of the phosphodiester groups in packaged DNA are
perturbed from that found in free DNA solutions, and the capsid
protein structures of mature virions are indistinguishable from empty,
DNA-free, particles.

Finally, we note from Eqs. (\ref{e4}-\ref{e7}) that the osmotic
pressure $\pi_{\text{in}}$ within the capsid, the pressure
$P_{\text{ejection}}$ associated with the chemical potential of DNA
ejection, and the pressure $P_{\text{DNA}}$ imparted by the capsid on
the DNA are entirely different thermodynamic quantities. Importantly,
it must be appreciated that these pressures are all equilibrium
quantities, and they cannot be used to describe a non-equilibrium
situation.

\section{Can $P_{\text{DNA}}$ and $P_{\text{ejection}}$ play any role
  in DNA ejection dynamics?\label{sec3}}

In this section, we investigate whether $P_{\text{DNA}}$ and
$P_{\text{ejection}}$ play a role in DNA ejection. We first discuss
the case of $P_{\text{DNA}}$ within the context of the continuum
mechanics model; in order to do so we start by analyzing the stability
of the DNA within a phage capsid.

\subsection{Stability of the DNA within a phage capsid\label{sec2b1}}

As described in Secs. \ref{sec1b}, continuum mechanics model starts
with the assumption that the DNA, within a capsid made of a perfectly
rigid material, takes a toroidal spool configuration on a hexagonal
lattice at all stages of ejection. Given that the outer radius of the
spool is fixed by the capsid dimensions, all structural aspects of the
spool are then determined by the spacing of the hexagonal
lattice. Phages with internal cores constitute a special case: the
radius of the innermost ring of the spool is simply given by the
radius of the core. For a given ionic condition $\{n_{\text{s,in}}\}$
inside the capsid --- which in turn is fixed by the chemical potential
$\{\mu_{\text s}\}$ of the small solutes in the external buffer (for
the case of counterions, via the Donnan equilibrium) --- and DNA of
length $L$ within a capsid of volume $V$, the free energy of the
material inside the capsid is a function of $a_H$ alone, where $a_H$
is the hexagonal lattice spacing; i.e.,
$F_{\text{p,in}}(V,L)|_{\{\mu_{\text s}\}}\equiv F(a_H)|_{\{\mu_{\text
    s}\}}$. Note that the configurational entropy of the DNA is not a
meaningful quantity in the continuum mechanics model, hence in this
section $F_c$ only contains curvature (bending) energy of the DNA;
i.e., $F_c(V,L)|_{\{\mu_{\text s}\}}\equiv U_c(a_H)$.

Within the continuum mechanics model, one can argue that $a_H$ is
fixed, via a trade-off between $F^{\text{(int)}}(a_H)|_{\{\mu_{\text
    s}\}}$ and $U_c(a_H)$ [see Eq. (\ref{e0})]. The interaction free
energy can be lowered by increasing $a_H$. However, since the ``volume
constraint'' (meaning that the total available volume within the
capsid is $V$) must be obeyed, increasing $a_H$ will reduce the
minimum radius of the DNA spool, because within the inner part of the
spool the DNA will be much more tightly bent, which will increase its
bending energy. In this manner, a reduction in interaction free energy
increases curvature energy and {\it vice versa\/}. In other words the
trade-off between these two quantities dictates that:
\begin{eqnarray}
\frac{\partial F^{\text{(int)}}(a_H)|_{\{\mu_{s}\}}}{\partial
  a_H}+\frac{\partial U_c(a_H)}{\partial a_H}=0,
\label{e1}
\end{eqnarray}
which minimizes the free energy $F(a_H)|_{\{\mu_{s}\}}$.  We denote
the value of $a_H$, obtained from Eq. (\ref{e1}) by $a^{(0)}_H$, and
the corresponding minimum of the free energy of the inside content of
the capsid by $F^{(0)}_{\text{p,in}}|_{\{\mu_{s}\}}\equiv
F(a^{(0)}_H)|_{\{\mu_{s}\}}$. If it is further assumed that most of
the interaction free energy variation due to variations in
$a^{(0)}_H$, caused by a virtual uniform expansion of the capsid [as
in Eq. (\ref{e3a})], involves the DNA helix, then the osmotic pressure
within the capsid can also be defined as
\begin{eqnarray}
  \pi_{\text{in}}=-\frac{2}{3\sqrt3a^{(0)}_HL}\frac{\partial
    F^{\text{(int)}}(a^{(0)}_H)|_{\{\mu_{s}\}}}{\partial a^{(0)}_H},
\label{e2}
\end{eqnarray}
where the volume that the DNA toroid occupies within the hexagonal
lattice is given by $3\sqrt3\left[a^{(0)}_H\right]^2L/4$. This analysis for the
specific case of the phage T7 --- a phage with an internal
proteinaceous core --- can be found in Ref. \cite{theo1}.

\subsection{Can $P_{\text{\emph{DNA}}}$ transmit down the tail
  tube?\label{sec3a}}

The only way $P_{\text{DNA}}$ can play a role in DNA ejection is if it
can transmit from within the capsid, where the bulk of the DNA is
located, down the tail tube. To evaluate this idea, we begin with the
appreciation that $P_{\text{DNA}}$ favors a uniform expansion of the
capsid. However, since DNA is not a continuum material,
$P_{\text{DNA}}$ does not trivially transmit along the helical axis
into the tail tube. Nevertheless, when the DNA organization within the
capsid is averaged over many phage particles, the picture that emerges
is that pressure on the toroidal spool of DNA does translate into a
{\it compressive\/} force $f$ acting along the contour of the
DNA. Following Sec. \ref{sec2b1}, we start by analyzing this force
within the continuum mechanics model, and we then ask whether this
force can transmit down the tail tube.
\begin{figure}[h]
\begin{center}
\begin{minipage}{0.4\linewidth}
\includegraphics[width=\linewidth]{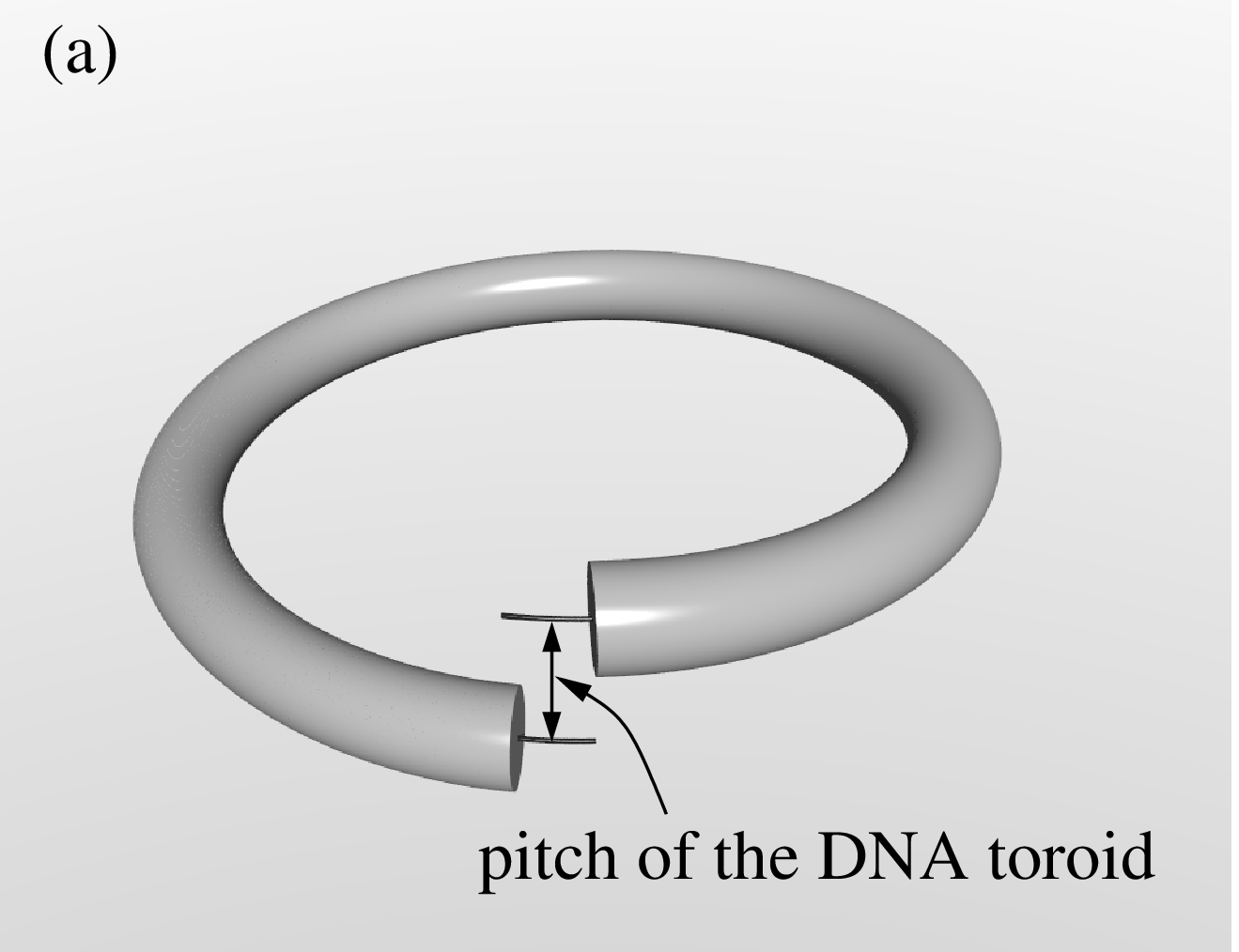}
\end{minipage}
\hspace{5mm}
\begin{minipage}{0.5\linewidth}
\includegraphics[width=\linewidth]{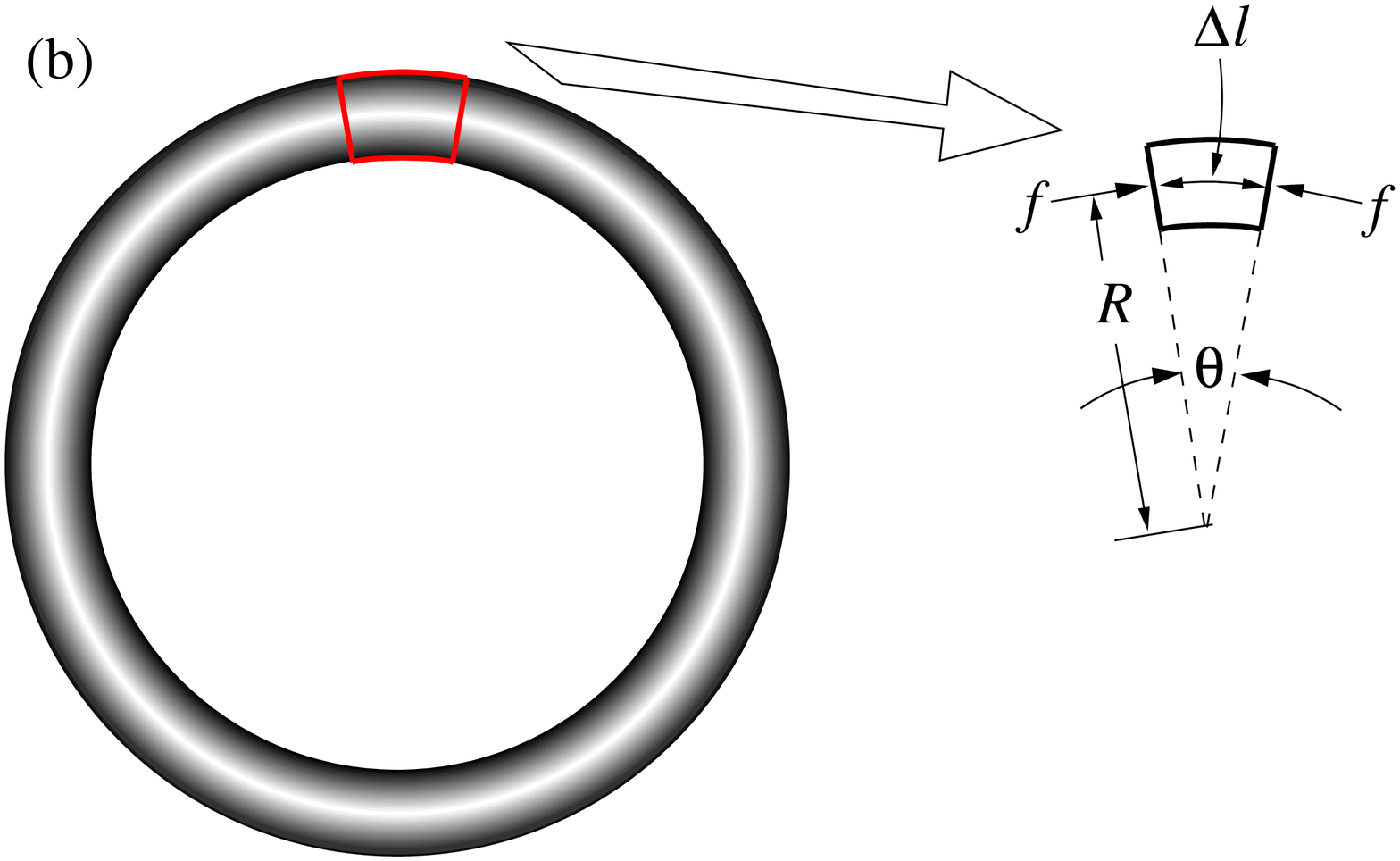}
\end{minipage}
\end{center}
\caption{(a) One full turn of the DNA toroid around the axis of the
  idealized toroidal spool in Fig. \ref{fig2}. (b) With the pitch
  ignored: the corresponding DNA ring of radius $R$, and the
  compressive force $f$ along the contour of the DNA, shown on a
  differential length element $\Delta l$ of the ring.\label{fig3}}
\end{figure}

We consider one full turn of DNA around the axis of the idealized
toroidal spool [Fig. \ref{fig3}(a)]. For simplicity, we ignore the
pitch of this turn, i.e., it is idealized as a DNA ring of radius $R$,
and we focus on a differential length segment $\Delta l$. The DNA ring
and the compressive force along the DNA contour are shown in
Fig. \ref{fig3}(b). The fact that the entire DNA spool is under a
pressure $P_{\text{DNA}}$ within the phage capsid implies that the
isolated ring itself is under a pressure $P$, but whose relation to
$P_{\text{DNA}}$ we have not specified.

Within the scheme of continuum mechanics, the force balance equation
for this differential length segment along the transverse direction
(along a line joining the differential length element and the center
of the ring) is given by
\begin{eqnarray}
f\sin\theta\approx f\theta=P\Delta l\,t,
\label{e8}
\end{eqnarray}
where $t$ is the cross-sectional diameter of the DNA. Further, the
geometrical con\-sideration from Fig. \ref{fig3}(b) tells us that
\begin{eqnarray}
\sin\theta\approx\theta=\Delta l/R,
\label{e9}
\end{eqnarray}
i.e.,
\begin{eqnarray}
f=P\,R\,t.
\label{e10}
\end{eqnarray}
We now note that, just like in Eq. (\ref{e1}), the DNA ring stays in
its configuration by the balance of two competing thermodynamic
forces, one derived from its curvature energy $u_c$, the other from
the energy of self-interaction with nearby DNA strands. The
thermodynamic force related to curvature energy favors an increase of
the radius of the ring, while the thermodynamic force related to the
self-interaction free energy favors a decrease. The magnitude of this
force is given by
\begin{eqnarray}
(2\pi R)\,P\,t=-\frac{\partial u_c}{\partial
R}=\frac{2\pi(k_BT)l_p}{R^2},
\label{e11}
\end{eqnarray}
implying that
\begin{eqnarray}
f=P\,R\,t=\frac{(k_BT)l_p}{R^2}.
\label{e12}
\end{eqnarray}
At room temperature ($T=300$ K) $k_BT=4.14\times10^{-21}$ Joules. If
we take $l_p=50$ nm, the widely used value, then
\begin{eqnarray}
f=\frac{2.07\times10^{-28}}{R^2}\,\mbox{Nm}^2.
\label{e13}
\end{eqnarray}
Thus, depending on the radius of the DNA ring, the magnitude of the
compressive force $f$ along the contour of the DNA can be a few
pN. Equation (\ref{e13}) tells us that the pressure on this DNA ring
is only in terms of the stiffness of the DNA, as we argued in
Sec. \ref{sec2c}. 

How much of this force can transmit down the tail tube? We should
remember that force is a vector quantity. Although the above exercise
shows that there is a compressive force along the contour of the DNA
ring that make up the toroidal spool, it is necessary to follow the
actual turns of the DNA leading into the tail tube in order to
identify how much of the force $f$ is actually being
transmitted. However, the definition of persistence length $l_p$ in
polymer physics means that while a DNA molecule of length $\lesssim
l_p$ is semi-flexible, a DNA molecule with length $\gg l_p$ is simply
a flexible polymer. It is therefore impossible to transmit compressive
forces along the contour of a DNA that is $\gg50$ nm.  This
consideration alone ensures that any compressive force along the
contour of, e.g., an $\sim16\,\,\mu$m long $\lambda$ genome will not
transmit into the tail tube.  Therefore, compressive forces on the
packaged phage genomic DNA cannot be important for DNA ejection.

Our estimate of $f$ having a magnitude of a few pN is, at first blush,
in apparent contradiction with the experiment by Smith {\it et al.}
\cite{smith}. A deeper look however immediately reveals that
Eq. (\ref{e13}) and the forces exerted by the packaging motor are
actually compatible. Thermodynamically, we have identified four
different pressures within the capsid: pressure on the DNA
$P_{\text{DNA}}$, hydrostatic pressure $P_{\text{hydro,in}}$, the
pressure experienced by the capsid $P_{\text{capsid}}$, and the
pressure $P_{\text{ejection}}$ associated with the chemical potential
of DNA ejection. These are all different quantities. Smith {\it et
  al.} \cite{smith} measure $P_{\text{ejection}}$; they assumed that
the work done against $P_{\text{ejection}}$ by the packaging motor is
used to pressurize the DNA and is conserved in the capsid. However,
most of the work done against $P_{\text{ejection}}$ by the packaging
motor is expended in increasing the osmotic pressure within the
capsid.  This increase is due to the expulsion of water molecules out
of the capsid --- a reverse osmosis process --- in order to allow the
phage genome being packaged to condense, while only a small part of
the work is used to increase $P_{\text{DNA}}$ The energy expended
during water expulsion is not conserved by the packaged genome, which
explains our lower estimate of $P_{\text{DNA}}$ inside the capsid.  In
this context it is important to remember that osmotic pressure is only
a measure for the chemical potential of water: the fact that the
osmotic pressure inside the capsid is high does not imply that the DNA
is under enhanced pressure.

\subsection{Is $P_{\text{\emph{ejection}}}$ important for DNA ejection
  dynamics?\label{sec3b}}

Equation (\ref{e4}) shows that for a mature phage in a given buffer,
the tail plug feels a thermodynamic pressure $P_{\text{ejection}}$
from inside the capsid. Given that $P_{\text{ejection}}$ is a
thermodynamically derived quantity (i.e., derivative of
$F_{\text{p,in}}$ with respect to $L$), it is clear that if genome
ejection were a quasi-equilibrium process, it would indeed govern the
ejection dynamics. However, the underlying assumption of
quasi-equilibrium is that at all stages of ejection the DNA remaining
within the phage head stays in a toroidal loop, with an
ever-increasing spacing between the toroidal strands as ejection
proceeds. This is the major thesis of the continuum mechanics model,
which uses $P_{\text{ejection}}$ to explain the {\it in vitro\/}
experimental data using phage $\lambda$.

Whether this thesis and its underlying assumption are correct or not
lies in the question of time scales. If the time-scales associated
with the equilibration of DNA (in the form of a toroidal spool on a
hexagonal lattice) and the small solute molecules within the phage
capsid (maintained at a fixed chemical potential by the external
buffer), are much smaller than the ejection time, equilibrium
pressures and forces can be used to describe genome ejection. However,
recent experimental evidence with T5 clearly demonstrates that as the
DNA leaves the capsid, it goes through a series of phase transitions
\cite{lef2}. The cryo-electron microscopic images of T5 virions during
DNA ejection strongly suggest that the process is non-equilibrium,
which invalidates the use of $P_{\text{ejection}}$ to describe its
dynamics. Comparable experiments have not yet been conducted with
other phages, but there is no justifiable reason to assume that the
genome of other phages, in particular those lacking a defined internal
core structure, whose DNA is packaged to the same density, would
behave otherwise. We are not aware of experimental studies that probe
the equilibration time-scales of confined DNA, such as within a phage
capsid. However, the high velocity of phage DNA ejection measured {\it
  in vitro}: up to 75 kb/sec (T5) with long pauses between distinct
steps \cite{mang}, and up to 60 kb/sec ($\lambda$) in 10 mM Na$^+$
buffer or $\sim20$ kb/sec in 10 mM Mg$^{2+}$ \cite{gray1}, makes it
highly unlikely that these are quasi-equilibrium processes. These
experimental data suggest that $P_{\text{ejection}}$ cannot be used to
describe DNA ejection dynamics even {\it in vitro}.

\section{An alternative mechanism of DNA ejection from phage
  virions\label{sec4}}

Sections \ref{sec2} and \ref{sec3} illustrate the fundamental problems
with the idea that the thermo\-dynamically derived ejection pressure
$P_{\text{ejection}}$ (or the corresponding ejection force
$A_{\text{DNA}}P_{\text{ejection}}$) causes DNA ejection from
phages. There is therefore a clear need for a mechanism that can
explain the physics of DNA ejection, one that is consistent with both
{\it in vitro\/} and {\it in vivo\/} experimental data. We suggest
such a mechanism below.

\subsection{DNA ejection {\it in vitro\/}\label{sec4a}}

Of the two main approaches for DNA ejection: continuum mechanics and
coarse-grained molecular mechanics models, only the former addresses,
using $P_{\text{ejection}}$, the DNA ejection mechanism in a way that
can be compared to {\it in vitro\/} experimental data. In these
creative experiments, phages are immersed in a solution containing PEG
and/or DNA condensing agents, and DNA is ejected from the virion when
triggered by the receptor protein. The length of DNA that remains
within the capsid at the end of the ejection process depends on the
concentration of the PEG and/or DNA condensing agent.

With the caveats we have discussed concerning the validity of using
$P_{\text{ejection}}$ to describe phage DNA ejection, and remembering
that $P_{\text{ejection}}$ also involves fitted parameters, the
continuum mechanics theory does provide good agreement with {\it in
  vitro\/} genome ejection data for $\lambda$ and SPP1 (e.g., Refs.
\cite{evi1,gray3,evi2,sao,evi3}). However, the data obtained with just
two phage systems, along with the physics of the continuum mechanics
model, have often been extrapolated to all phages and in addition to
explain all phage DNA ejection {\it in vivo\/}. These generalizations
ignore data from the third phage system that has been studied {\it in
  vitro}, which provides rather different conclusions. At the presumed
pressures internal to the largely full T5 capsid, {\it in vitro\/}
experimental data can be fitted to the continuum mechanics model, but
at low to moderate pressures, when approximately half or less of the
genome remain in the capsid, multiple populations are found to
co-exist \cite{lef1}: some phages have completely ejected their
genomes, whereas others have ejected a varying amount of DNA that is
not dependent on the external osmotic pressure. Furthermore, at $\le2$
atm external osmotic pressure, most T5 virions completely eject their
DNA. In contrast, the same external pressure prevents ejection of
$\sim40\%$ of the $\lambda$ genome \cite{gray3}. Theories dependent on
$P_{\text{ejection}}$ are clearly unable to explain how {\it all\/}
phages eject their DNA, even {\it in vitro}.

\subsubsection{Hydrodynamic model of phage DNA ejection\label{sec4a1}}

As discussed in Sec. \ref{sec3b}, in an {\it in vitro\/} experiment
when a mature phage is equilibrated in a certain buffer solution, the
difference between $\pi_{\text{in}}$ and the osmotic pressure
$\pi_{\text{out}}$ of the buffer solution determines the thermodynamic
incentive for water molecules to enter the capsid. Equation (\ref{e6})
shows that if the phage tail is plugged, this incentive is
counter-balanced by a hydrostatic pressure gradient across the capsid
shell (and the capsid, being of high but not infinite rigidity, will
be slightly expanded from its relaxed state under the enhanced
hydrostatic pressure from inside). We propose that it is this
thermodynamic incentive of water molecules to enter the capsid that
flushes the DNA out when the tail plug is opened by the action of the
appropriate receptor. We refer to this idea as the hydrodynamic model
of DNA ejection.

When the tail plug is opened by the appropriate receptor, e.g., LamB
protein in the case of $\lambda$, FhuA for T5, or YueB780 for SPP1, it
is likely that the excess hydrostatic pressure within the capsid is
partially relieved by a transient water flow down the tail tube and,
and as a result, the capsid will reduce in size. Although water flow
down the tail-tube will exert a hydrodynamic force that is likely to
drag along the DNA, one end of which is already inserted into the
tail-tube, this process will not last long enough to drag the entire
genome out of the virion. While a very small contraction of the capsid
is unlikely to be detected even by state-of-the-art experiments, we
can provide a ``guesstimate'' for how much DNA could be dragged out by
this transient water release. With a typical capsid diameter $\sim60$
nm (assumed spherical), let the capsid diameter contract by $0.1$ nm;
the water released is $4\pi(30)^2(0.05)$ nm$^3$. With a typical tail
tube inner diameter $\sim4$ nm, its cross-sectional area is
$~\pi(2^2)$ nm$^2$ (ignoring the cross-section of the DNA within the
tail tube) and the length of the water column passing into the
tail-tube $=4\pi(30^2)(0.05)/[\pi(2^2)]$ nm $=45$ nm. Assuming the DNA
flows freely with this water column down the tail, in this example the
DNA, which occupies 25\% of the cross-sectional area of the tube, can
be moved 60 nm, corresponding to $\sim180$ base pair of a phage
genome. 

\begin{figure}[h]
\begin{center}
\includegraphics[width=\linewidth]{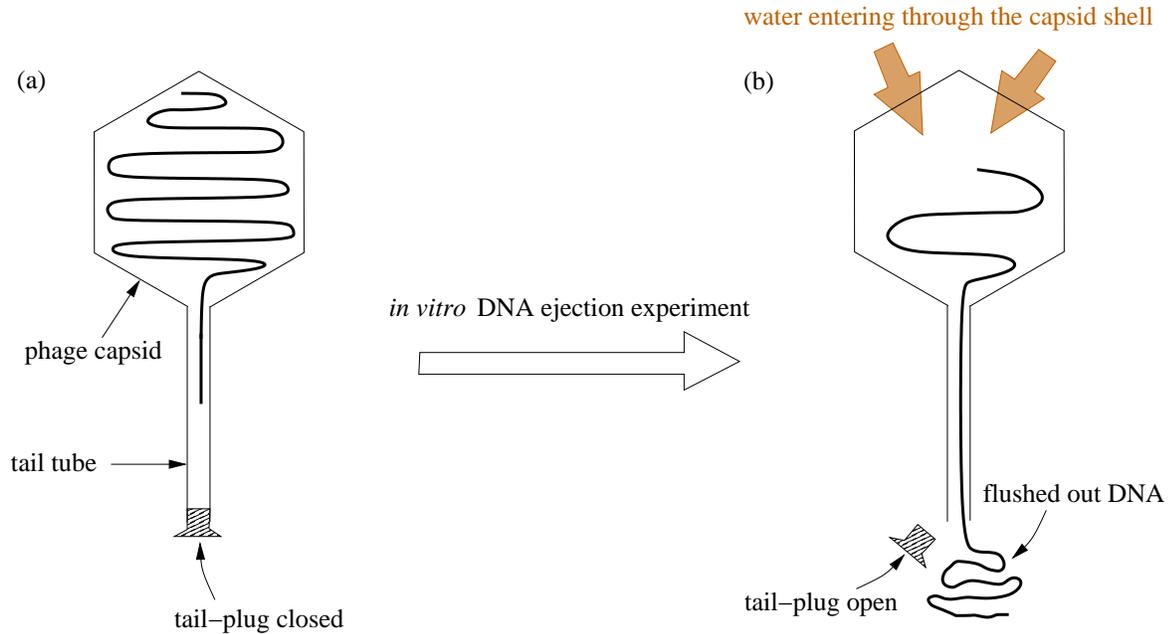}
\end{center}
\caption{Schematic of an {\it in vitro\/} DNA ejection experiment. (a)
  Mature phage equilibrated in a certain buffer solution. The
  difference between $\pi_{\text{in}}$, the pressure on the DNA, and
  the osmotic pressure $\pi_{\text{out}}$ of the buffer solution is
  counter-balanced by a hydrostatic pressure difference across the
  capsid shell [Eq. (\ref{e6})]. (b) The tail plug is opened by the
  action of the receptor, following any transient water release down
  the tail-tube (not shown), water enters the phage capsid and flushes
  the DNA into the buffer.\label{fig4}}
\end{figure}

Once the water pressure difference across the capsid shell is relieved
by this transient water flow, so long as the difference between the
osmotic pressure of the remaining DNA within the capsid and
$\pi_{\text{out}}$ stays positive, water will have a thermodynamic
incentive to move into the capsid using whichever path it can
find. However, since the capsid has a finite volume, any water
movement into the capsid can only occur at the expense of an equal
volume of water, now containing DNA (as a solute) being ejected
through the tail tube. In other words, following the initial transient
water release down the tail-tube, continued water movement from the
buffer into the capsid up an osmotic gradient will cause the DNA to be
ejected. This process will stop when the osmotic pressure of the
leftover amount of DNA in the capsid equals $\pi_{\text{out}}$. The
physics behind this {\it in vitro\/} DNA ejection experiment (omitting
the transient water flow down the tail tube due to the excess
hydrostatic pressure) is schematically shown in Fig. \ref{fig4}.

The hydrodynamic mechanism for DNA ejection commensurate with {\it
  in vitro\/} experimental data. All that is required to drive DNA
ejection is an incentive for the water molecules to enter the capsid,
which is decided dynamically in a non-equilibrium manner.  There is no
{\it a priori\/} assumption that phage genome ejection is a
quasi-equilibrium process, and thus the model is not dependent on
thermodynamic analyses.

The difference between the osmotic pressure of the residual DNA in the
capsid and $\pi_{\text{out}}$ is controlled by two aspects: (i) the
presence of any small DNA condensing agents that reduces the osmotic
pressure of the capsid content without altering $\pi_{\text{out}}$,
and (ii) the presence in the external buffer of large solute molecules
like PEG that increase $\pi_{\text{out}}$, but do not affect the
inside osmotic pressure as they cannot penetrate into the capsid. Both
achieve the same thermodynamic incentive differential for water
between inside and outside the capsid, and are qualitatively in
agreement with {\it in vitro\/} DNA ejection data
\cite{evi1,gray3,evi2,jeem1,lef1,def,mang,gray1,rick,sao,evi3}. When
the osmotic pressure of any DNA remaining in the capsid
$=\pi_{\text{out}}$ (such as at the end of the ejection process),
there will be no further flow of water into the capsid, and thus no
more DNA ejection.

The hydrodynamic model requires that the time-scale of water flow into
the capsid be sufficiently fast to match the time-scales for DNA
ejection {\it in vitro}. To this end, we refer to an experiment where
a mature wild-type $\lambda$ capsid was locally deformed (compressed)
by up to 25\% of its original volume with an AFM tip
\cite{iva1}. Compression of the capsid without it rupturing has to
drive out internal water and small solutes. When the AFM tip was
removed, the virion capsid returned to its normal size and shape
within $\sim4$ ms. This suggests that the large surface area to volume
ratio of the capsid allows copious amounts of water to diffuse --- a
slow process relative to hydrodynamic processes --- across the capsid
shell in a very short time.  Extrapolating these data to yield a
complete exchange of the contents of the capsid with the external
buffer gives 20 ms; assuming that DNA moves freely, this value results
in a rate of phage genome ejection {\it in vitro} at least 50 times
the maximum measured to date.

By itself, the hydrodynamic model cannot explain either the stepwise
ejection of DNA from T5 particles or the coexistence of virions
containing different amounts of DNA at low external pressures.  A
partial {\it ad hoc\/} solution for T5 (one that is also necessary for
the $P_{\text{ejection}}$ model of DNA ejection) is that the tail tube
becomes temporarily blocked by protein conformational changes, thereby
impeding water flow --- and thus stopping DNA ejection. Note that water
flow from the external buffer into the capsid interior will be
simultaneously impeded.  Release of the blockage by a further
conformational change in the tail proteins then allows resumption of
water flow and thus of DNA ejection.  Interestingly, the apparent
activation energy for releasing the block in the step-wise ejection
process of T5 DNA ejection has been experimentally shown to be
independent of the amount of DNA remaining in the capsid
\cite{gray3,raspaudnew}.

In summary, the hydrodynamic model is fully compatible with the
experimental data for $\lambda$ and for SPP1 DNA ejection {\it in
  vitro}. Like thermodynamics-based models (e.g., continuum mechanics
or molecular dynamics), the hydrodynamic model requires an {\it ad
  hoc\/} assumption to explain the step-wise process of T5 DNA
ejection.  However, the hydrodynamic model is not dependent on the
critical assumption --- used in $P_{\text{ejection}}$-based models ---
that the encapsidated DNA remains at equilibrium at all times during
the ejection process.  Furthermore, as we argue below, only the
hydrodynamics model explains how complete genome ejection can be
achieved in the face of an opposing force.  This is the situation in
natural infections of bacterial cells.

\subsection{DNA ejection {\it in vivo}\label{sec4b}}

Before discussing specific mechanisms of phage DNA ejection {\it in
  vivo}, it is instructive to consider some salient points about phage
infections.

\subsubsection{Phage infection of bacteria\label{sec4b1}}

Largely because of influential textbooks and, until recently, of only
little experimental information, it is not widely appreciated that
{\it all} phages eject proteins into the cell \cite{mol1,mol2}.  At a
minimum, the protein(s) comprising the tail plug must be removed, and
for long-tailed phages (like T4, T5, SPP1 and $\lambda$) the
tapemeasure protein, which determines the precise length of the tail,
must be ejected in order to allow the phage genome to pass through the
tail tube.  Short tailed phages may eject internal proteins into the
cell to extend --- at least functionally --- their tail so that it can
span the infected cell envelope \cite{mol3}.  

Some phages eject many different protein molecules from their capsid,
some necessarily before, others perhaps after, genome ejection. To
give just two examples, T7 virions ejects $\sim75$ molecules
representing five different protein species into the cell prior to DNA
penetration of the cell cytoplasm \cite{kemp2}, and T4 virions eject
$\sim1000$ IP (internal protein) molecules into the cell
\cite{onorato}. Clearly, any model purporting to explain even dsDNA
phage genome ejection {\it in vivo\/} must also accommodate ejection
of virion proteins.

Phage virions can also eject more than one DNA molecule
\cite{leffers,coren}. However, as the diameter of the channel through
the portal complex and tail tube (inner diameter $\sim40$ \AA) cannot
accommodate more than a single DNA helix, the leading end of the
second (and perhaps subsequent molecules) must find the exit channel
without a built-in guiding vectorial force. Genome ejection by
single-stranded DNA (ssDNA) or RNA viruses, whose nucleic acid is
packaged at much lower densities than dsDNA phages, should also be
explained by models describing phage DNA ejection. Packaged genomes of
these phages are not thought to be under pressure, and their mode of
genome ejection has therefore not been considered by continuum
mechanics or molecular mechanics models.

The {\it E. coli\/} cytoplasm has a positive osmotic pressure of
several atm above the environment (under various growth conditions the
pressure has been estimated to vary between 2 and 15 atm, with 3.5-5
atm being commonly accepted values \cite{stock,koch1}). This osmotic
pressure gradient is counter-balanced by a hydrostatic pressure
differential (turgor) \cite{koch} that enables the cell to enlarge
during growth \cite{koch2}. Gram-positive cells, such as {\it Bacillus
  subtilis}, the host for phage SPP1, have much higher turgor:
$\sim19$ atm \cite{what}. If turgor is assumed to provide an opposing
force to phage genome ejection into the cytoplasm, then a
$P_{\text{ejection}}$-based mechanism cannot possibly explain complete
genome ejection into cells. This problem was recognized in the first
osmotic suppression of genome ejection experiments {\it in vitro\/}
\cite{evi2}, but has often been ignored in subsequent general
statements about how phages infect cells {\it in vivo}. Interestingly,
the osmotic pressure of the cytoplasm declines as bacteria enter
stationary phase, which should allow better genome penetration by a
$P_{\text{ejection}}$-based ejection mechanism. However, stationary
phase bacteria are not readily infected by most phages
\cite{schrader}, which infect exponentially growing cells, with their
higher turgor pressure, much more efficiently.

A bacterial cytoplasmic membrane is composed of a phospholipid bilayer
that is impermeable to most molecules other than water and
glycerol. An electrical potential is maintained across this
membrane. A strong osmotic gradient also exists between the external
medium and the cell cytoplasm. The higher internal pressure (turgor)
is necessary for bacteria to enlarge and undergo cell division. In
addition, at all times, a K$^+$ concentration gradient [K$^+$], with
[K$^+$] high inside, is maintained in cells (a reverse gradient of
Na$^+$ concentration also usually exists). These conditions are
perturbed when a bacterium is infected by a phage as a direct
connection between the cytoplasm and the external medium is opened.
This connection passes through the phage capsid and tail, and if it is
open, and, as we have argued, water will flow from the external medium
into the cell to neutralize the overall osmotic
gradient. Simultaneously, K$^+$ will flow from the cytoplasm into the
external medium. Furthermore, the membrane potential can no longer be
maintained.

Advocates of $P_{\text{ejection}}$-based genome ejection {\it in
  vivo\/} have made various {\it ad hoc\/} suggestions to explain how
an entire phage genome can enter the cell cytoplasm, including the
involvement of cytoplasmic, non-sequence-specific, DNA-binding
proteins or condensation of the phage DNA in the crowded cell
cytoplasm. There is however, no direct experimental support for any of
these ideas. Condensing the entering phage genome in the bacterial
cytoplasm would effectively prevent its transcription, but efficient
--- and immediate --- gene expression is precisely the primary
strategy of phage infections. Furthermore, with the specific exception
of second-step transfer of T5 DNA, it is hard to imagine how entering
phage DNA could effectively compete with the 100-fold higher DNA
concentration of the bacterial chromosome for those proteins in order
to complete genome internalization in a kinetically reasonable time
frame. T5 second-step transfer requires the prior synthesis of two
T5-encoded DNA-binding proteins, and early T5-encoded nucleases
completely degrade the bacterial chromosome \cite{McCorq}, thereby
removing competitor DNA. Nevertheless, sequence-specific DNA-binding
proteins, which have no or fewer binding sites on bulk chromosomal
DNA, and can therefore effectively bind to incoming phage DNA {\it in
  vivo}, have been shown to catalyze genome internalization of phage
T7 and its relatives \cite{mol1}, and also of N4 \cite{choi}.

\subsubsection{The hydrodynamics model is consistent with the physiology
  of phage infection\label{sec4b2}}

DNA ejection {\it in vivo\/} and {\it in vitro\/} are actually quite
different. The major distinction is that when the tail is unplugged
{\it in vitro}, the environment at the distal end of the tail tube is
the same buffer that surrounds the capsid.  In contrast, during a
natural infection the distal end of the tail tube is in the cytoplasm
of the infected cell, while the environment of the capsid is the
growth medium. Consequently, the osmotic pressure of the medium that
surrounds the phage capsid is different from that at the opening of
the tail tip. 

\begin{figure}[h]
\begin{center}
\includegraphics[width=\linewidth]{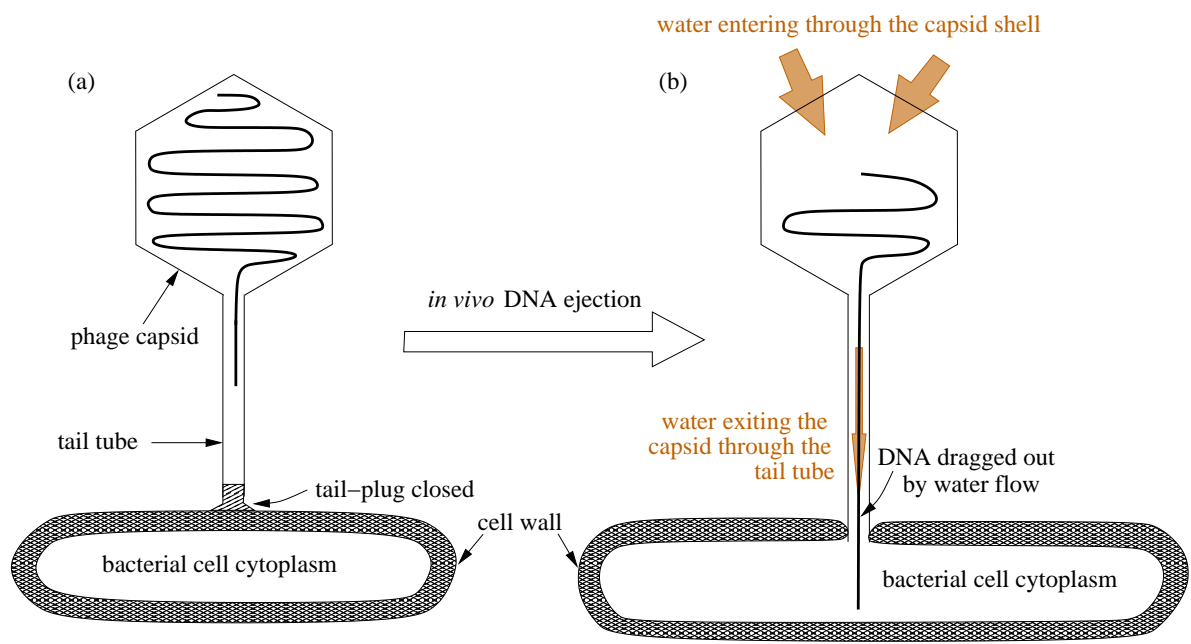}
\end{center}
\caption{Schematic of DNA ejection {\it in vivo}. (a) Mature phage
  adsorption on a bacterial cell. (b) The plug blocking the tail is
  opened or removed as the cell envelope is breached. Water and small
  solutes enter the phage capsid from the culture medium and drains
  through the tail tube into the cytoplasm of the infected cell,
  dragging the DNA along with it.\label{fig5}}
\end{figure}

We denote the osmotic pressure of the culture medium by
$\pi_{\text{out}}$, and that of the cytoplasm of the infected cell by
$\pi_{\text{cyt}}$; note that $\pi_{\text{cyt}} > \pi_{\text{out}}$ as
the bacteria need to maintain positive turgor in order to grow
\cite{koch1}. Although in $P_{\text{ejection}}$-based model of phage
genome ejection turgor is assumed to provide a resisting force against
$P_{\text{ejection}}$, it is the bacterial Achilles heel that actually
promotes complete genome ejection by a hydrodynamics mechanism. When
the tail plug is removed by interaction with its receptor, water will
flow up the osmotic gradient --- one that exists between the growth
medium and the cell cytoplasm --- from the growth medium, through the
capsid shell and tail tube into the cell cytoplasm. This directional
water flow can drag the DNA into the infected cell. This mechanism is
schematically shown in Fig. \ref{fig5}. It should be noted that
bacterial cells respond immediately to changes in $\pi_{\text{cyt}}$
and they will always maintain some turgor even during phage infection.

When bacteria enter stationary phase, their turgor pressure drops
considerably. This reduces the osmotic gradient between the external
medium and the cytoplasm, in turn this reduces the strength of water
flow from the culture medium into the cytoplasm. This may explain why
bacteria in stationary phase are not easily infected by most
phages. Notably, T7 is an exception, infecting stationary (and
starved) cells at normal efficiency \cite{schrader}, even though
subsequent phage development may be compromised by a lack of
biosynthetic capacity in the cell.

Following infection by all but one of the phages we are aware of that
have been tested, there is a transient drop in membrane potential and
a transient release of intra-cytoplasmic ions into the external medium
\cite{kala,drei}. Both occur during the period that DNA is thought to
be ejected from the virion into the cell cytoplasm.  In order to
evaluate the role of the membrane potential during T4 infection, an
extensive series of electrochemical studies that monitored ion fluxes
were conducted (Reviewed in Refs. \cite{grin1,gold1}). Experiments were
often combined with infective center assays after Hershey-Chase
blending of phage-cell complexes, allowing estimates of $10^3$-$10^4$
bp/sec for the rate of T4 genome internalization in vivo
\cite{kala,grin1,grin2,gold1}.  Furthermore, T5 releases its DNA in
two distinct steps \cite{McCorq}. During each step, cytoplasmic K$^+$
leaks into the external medium, while during the pause between the
steps no leakage occurs \cite{boul}. Thus, ion leakage is associated
with phage DNA ejection, and if ions are leaking out, water should be
flowing from the external medium through the phage virion into the
cell cytoplasm to reduce the osmotic gradient.  This is exactly what
is expected if hydrodynamic forces are dragging the phage genome into
the cell.

There are conflicting published data on membrane depolarization and
ion leakage for phage T7, but it is now clear that T7 infection does
not result in either phenomenon \cite{kuhn,ram}, indicating that no
open channel that could allow water flow exists between the
environment and the infected cell cytoplasm. These are significant
observations, because, unlike other phages, the internalization of T7
and related phage genomes is entirely catalyzed by molecular
motors. These motors are enzymes that utilize cellular energy to
transport the entire genome into the cell, each functioning at a
constant rate regardless of how much DNA remains in the phage head
\cite{kemp,gar1,gar2,gar3,strut,chang1}. Thus, in the absence of
hydrodynamic forces, energy-requiring enzymes are necessary to effect
phage genome internalization by the cell.

The hydrodynamic model of DNA ejection does not require secondary {\it
  ad hoc\/} mechanisms to complete genome transfer into the cell.
Because the water flow is determined by the values of
$\pi_{\text{cyt}}$ and $\pi_{\text{out}}$, the ejection process is
(relatively) independent of DNA condensing agents. A bacterial cell
responds immediately to changes in $\pi_{\text{cyt}}$ and will
maintain some turgor even during phage infection. Thus, hydrodynamic
flow from the external environment through the phage particle into the
cell cytoplasm will continue until genome internalization is complete
and the channel is closed.  (It is not known how this occurs in any
phage system, but all models of phage DNA ejection must invoke this
step in order to prevent a permanent loss of the cellular membrane
potential. If the potential is totally collapsed, the cell would die
and no phage progeny would ever result.) Furthermore, although low
concentrations of DNA condensing agents have a major effect on
$\pi_{\text{in}}$, they have little effect on either
$\pi_{\text{out}}$ or $\pi_{\text{cyt}}$ (e.g., intracellular levels
of polyamines, which are critical for ribosome and chromosome
stability are tightly regulated). This means that phage DNA ejection
{\it in vivo\/} should be relatively independent of the presence of
DNA condensing agents in the growth medium. Indeed, ongoing
experiments in the laboratory of one author (IJM) suggest that the
latent periods and burst sizes of various phages (including parallel
experiments with $\lambda$ and the deletion mutant $\lambda$b221, the
latter having a substantially reduced value, relative to wild-type
$\lambda$, of $\pi_{\text{in}}$ ) are not affected by the presence of
1 mM spermine in the bacterial growth medium.

The hydrodynamic model also provides a mechanism for the ejection of
single-stranded genomes and proteins into the cell during infection.
This is important as all phages eject protein molecules into the
infected cell, and a significant fraction of phages do not contain
dsDNA packaged at the high density of $\sim500$ mg/ml.  Any molecule
in the path of water flowing along the osmotic gradient between the
external medium and the cell cytoplasm can be driven down the tail
tube and into the cell.  As this water flow is necessarily vectorial,
nucleic acids or proteins to be ejected do not need to be positioned
in the tail tube in the mature phage particle.  Neither do they need
to possess internal energy to drive their ejection from the capsid.

In summary, with the caveat that, to explain the multi-step process of
T5 and perhaps other phages ($\phi29$ is one known example
\cite{gon}), an additional {\it ad hoc\/} mechanism of protein
conformational changes that temporarily block water flow from the
external fluid into the cell cytoplasm, in turn temporarily stopping
DNA ejection, the hydrodynamic model of phage DNA ejection is
consistent with --- and importantly can explain --- all the
observations we are aware of that have been made using any phage-host
combination {\it in vivo}.

However, theories need to be tested experimentally.  Determining the
kinetics of phage genome internalization into infected cells would go
a long way to supporting or refuting competing models.  The internal
capsid pressure models predict that the late stages of genome ejection
should slow towards a zero rate as the residual pressure equals that
in the cell cytoplasm, and that a secondary process using, for
example, polyamines or DNA-binding proteins to interact with the
entering the DNA completes the process.  A secondary process is
likely to occur with different kinetic parameters and may be
experimentally detectable.  Conversely, the hydrodynamic model would
predict a largely constant rate of genome internalization, as the
osmotic pressure gradient between the cell cytoplasm and the
environment is unlikely to change substantially while viral DNA enters
the cell.  Unfortunately, it has proved very hard to measure directly
the {\it in vivo\/} kinetics of genome entry outside the T7 family of
phages, although some measurements have been made with phage N4
\cite{demi}, and experiments with both $\lambda$ and T5 are being
initiated in the Molineux laboratory.  However, it may be possible,
using a combination of polyamines and external osmolytes, to increase
the cytoplasmic osmotic pressure above that inside the mature phage
head.  If phage infection still occurs normally under those
conditions, it would then provide irrefutable evidence that internal
capsid pressure is not important for infection, although it would not
provide any direct evidence for the hydrodynamic model.

\section*{Acknowledgements}

Work in the laboratory of I. J. M. has been supported by grant GM32095
from the National Institutes of Health (USA). The authors are grateful
to Anton Petrov and Steve Harvey for their extensive and constructive
critiques of the paper.  We also thank Theo Odijk for discussions, and
Marco Bosch for help with Fig. \ref{fig3}.

\end{document}